\documentstyle[epsf,12pt]{article}

\def\ket#1{| #1 \rangle}

\def\braketb#1#2{\langle #1 | #2 \rangle}
\def\braketc#1#2#3{\langle #1 | #2 | #3 \rangle}
\def\braketr#1#2#3{\langle #1 || #2 || #3 \rangle}

\def\dket#1{| #1 \rangle\!\rangle}

\def\dbraketc#1#2#3{\langle\!\langle #1 | #2 | #3 \rangle\!\rangle}

\def\gtsim{\mathrel{\hbox{\raise0.2ex
     \hbox{$>$}\kern-0.75em\raise-0.9ex\hbox{$\sim$}}}}
\def\ltsim{\mathrel{\hbox{\raise0.2ex
     \hbox{$<$}\kern-0.75em\raise-0.9ex\hbox{$\sim$}}}}
\def\gtlt{\mathrel{\hbox{\raise0.7ex
     \hbox{$>$}\kern-0.75em\raise-0.3ex\hbox{$<$}}}}
\def\ltgt{\mathrel{\hbox{\raise0.7ex
     \hbox{$<$}\kern-0.75em\raise-0.3ex\hbox{$>$}}}}
\def\gele{\mathrel{\vbox{\hbox{\raise1.0ex
     \hbox{$\ge$}\kern-0.75em\raise-0.7ex\hbox{$<$}}}}}
\def\lege{\mathrel{\vbox{\hbox{\raise1.0ex
     \hbox{$\le$}\kern-0.75em\raise-0.7ex\hbox{$>$}}}}}

\def\dI{ {\mit\Delta I} }
\def\Ii{ I_{\rm i} }
\def\If{ I_{\rm f} }

\def\dK{ {\mit\Delta K} }
\def\Ki{ K_{\rm i} }
\def\Kf{ K_{\rm f} }
\def\cJ{ {\cal J} }
\def\cD{ {\cal D} }

\def\tQ{ {\widetilde Q} }

\def\i{ {\rm i} }
\def\f{ {\rm f} }
\def\rot{ {\rm rot} }

\begin{document}

\begin{center}
{\large\bf Heuristic quantization of the cranking model}
\vspace{0.07cm}

{Takashi Nakatsukasa\footnote{
Present address : Department of Physics, UMIST, P.O.Box 88, Manchester,
M60 1QD, U.K.}
and Yoshifumi R. Shimizu$^*$}

{\it AECL, Chalk River Laboratories
Chalk River, Ontario K0J~1J0, Canada\\
$^*$Department of Physics, Kyushu University, Fukuoka 812, Japan}
\end{center}

The cranking model, which is able to describe nonlinear effects of
nuclear rotation from a microscopic view point,
has been extensively applied to studies of rapidly
rotating nuclei.
However, the semi-classical treatment of the angular momentum
prevents the model from being directly applied to the study of
the electromagnetic transition probabilities at low spin.
The unified model has been a basic tool to analyze
the electromagnetic transition rates of rotational bands.
In this model, although the nuclear rotation is described by
macroscopic (``rotor'') wave functions,
the quantum-mechanical angular-momentum algebra is precisely
taken into account.
Thus, these two models, the cranking and unified models, have
complementary merits and demerits.
If we can combine the merits of both models,
we may be able to obtain a feasible method of microscopically calculating
transition probabilities of rotational bands.
In this paper, we introduce a simple ``quantization'' procedure
for the cranking model and discuss
Coriolis coupling effects on
the $I^\pi = 3^-$ octupole states as an example of its application.
See Ref.~[1] for details.

Since the {\it quantum} angular momentum
becomes approximately {\it classical} in the high-spin limit,
the transition probabilities can be estimated at this limit.
Marshalek derived
a simple formula for the reduced transition amplitude
in the leading order of $1/I$~[2];
\begin{equation}
   \frac{\braketr{\If}{Q_\lambda}{\Ii}}{\sqrt{2\Ii + 1}} \approx
   \braketc{\f}{\tQ_{\lambda \mu=\dI}}{\i} \ ,
\label{Marshalek}
\end{equation}
where $\dI = \If - \Ii$ and $\tQ_{\lambda \mu}$ is the transition
operator defined with respect to the cranking (rotation) axis,
$ \tQ_{\lambda \mu}
  = i^{-\mu} \sum_{\nu} d^\lambda_{\mu,\nu} \left(-\pi /2 \right)
    Q_{\lambda \nu}$
(the operator defined with respect to the symmetry
axis is denoted as $Q_{\lambda \nu}$ without a tilde).
The initial and final states in the cranking model,
$\ket{\i}$ and $\ket{\f}$, have good signature-quantum numbers
and are constructed as a linear combination of $\dket{\pm K_{\i (\f)}}$
which have good $K$-quantum numbers in the limit of $\omega_\rot=0$.
One may understand that
the direction of the angular momentum exactly coincides
with the cranking axis in this formula,
which is valid only in the high-spin limit.

If we simply apply the formula (\ref{Marshalek}) to the low-spin limit
($\omega_\rot =0$),
we obtain for the $K$-allowed transitions
within the first order with respect to $\omega_\rot$,
\begin{equation}
  \frac{\braketr{\If}{Q_\lambda}{\Ii}}{\sqrt{2\Ii + 1}}
   = C_{\i\f} \Biggl(
  \Bigl[ \dbraketc{K_\f}{Q_{\lambda \dK}}{K_\i} \Bigr]_0
         d^\lambda_{\dI, \dK} \Bigl(-\frac{\pi}{2}\Bigr)
    + \sum_{\rho=\pm 1}
  \Bigl[ \frac{d\dbraketc{K_\f}{Q_{\lambda, \dK+\rho}}{K_\i}}
              {d\omega_\rot} \Bigr]_0
  \omega_\rot \, d^\lambda_{\dI, \dK+\rho} \Bigl(-\frac{\pi}{2}\Bigr)
  \Biggr)\ ,
\label{lowspin}
\end{equation}
where $\dK=\Kf-\Ki$ and
the overall factor $i^{-\dI}$ and the signature-dependent terms like
$\dbraketc{K_\f}{Q_{\lambda \nu}}{-K_\i}$
were omitted in the r.h.s. for simplicity.
$C_{\i\f}=1$ if $\Ki=\Kf=0$ or $\Ki\neq 0$, $\Kf\neq 0$,
otherwise $C_{\i\f}=\sqrt{2}$.
Here $[*]_0$ means that the expression is evaluated
by taking the limit $\omega_\rot \to 0$.
Since the $d$ function $d^\lambda_{\dI, \dK}(-\pi/2)$
accounts for only the classical geometry of angular momentum
(the argument $-\pi/2$ means the direction of the angular momentum
vector is perpendicular to the symmetry axis),
Eq.(\ref{lowspin}) cannot describe
the correct intensity relations at low spin.
We need to quantize this equation.

In order to quantize the semi-classical angular momentum,
again we consider the high-spin limit where quantum
effects become less important.
In the limit of $\Ii, \If \gg \Ki, \Kf$,
\begin{equation}
     d^\lambda_{\dI,\dK}(-\frac{\pi}{2}) \approx
          \braketb{\Ii \Ki \lambda \dK}{\If \Kf} \ ,
     \quad\quad \mbox{with}\quad \dI=\If-\Ii \mbox{ and }\dK=\Kf-\Ki \ .
\label{d_CG}
\end{equation}
Thus, if we replace the $d$ function in the leading-order term of
Eq.(\ref{lowspin}) by the equivalent Clebsch-Gordan coefficient
(\ref{d_CG}),
we obtain the same structure as that of the leading-order intensity
relation in the unified model,
$C_{\i\f} \dbraketc{K_\f}{Q_{\lambda \dK}}{K_\i}
 \braketb{\Ii \Ki \lambda \dK}{\If \Kf}$.
This replacement allows us to apply this formula at low spin
because the C.-G. coefficient
$\braketb{\Ii \Ki \lambda \dK}{\If \Kf}$ incorporates the
quantum angular-momentum algebra.

In order to generalize this replacement to higher-orders,
it is convenient to rewrite Eq.(\ref{d_CG}) as
\begin{equation}
   d^\lambda_{\dI,\dK} \Bigl(-\frac{\pi}{2} \Bigr)
       \,\rightarrow\,\,
 \frac{\braketr{\Kf\If}{\cD^\lambda_{*, \dK}}{\Ki\Ii}_{\rm unsym}}
           {\sqrt{2\Ii + 1}} \ .
\label{quantization_1}
\end{equation}
This means that
the $d$ function in Eq.(\ref{lowspin}) corresponds to the reduced
matrix element of the $\cD$ operator with respect to the rotor part of
the unsymmetrized wave functions in the unified model.
Roughly speaking, this may be seen as a ``quantization'';
$d$ function $\rightarrow$ $\cD$ operator.
Using the relation $\omega_\rot \approx \langle J_x \rangle /\cJ
\rightarrow I_\pm /\cJ$,
now we may introduce a ``quantization'' rule for the first-order
correction terms (the second terms in the r.h.s. of Eq.(\ref{lowspin}));
\begin{equation}
   \omega_\rot \, d^\lambda_{\dI, \dK \pm 1} \Bigl(-\frac{\pi}{2} \Bigr)
       \,\rightarrow\,\,
      \frac{1}{\cJ}
      \frac{\braketr{\Kf\If}
          {\,{\displaystyle \frac{1}{2} }
      \{I_{\pm}, \cD^\lambda_{*, \dK \pm 1}\}\,}{\Ki\Ii}_{\rm unsym}}
           {\sqrt{2\Ii + 1}} \ ,
\label{quantization_2}
\end{equation}
where $\cJ$ is the moment of inertia of the band under consideration
at $\omega_\rot=0$.
The appropriate operator ordering in the r.h.s.
is discussed in Ref.~[1].
These higher-order terms take account of the Coriolis-coupling effects.
It is worth noting that,
with $\omega_\rot = I/\cJ$,
the the l.h.s. and r.h.s. of
Eqs.(\ref{quantization_1}, \ref{quantization_2}) become
exactly equivalent in the high-spin limit,
because the {\it quantum} values (r.h.s.)
are approaching the {\it classical} values (l.h.s.)
at this limit.

Using these ``quantization'' rules,
one can obtain the explicit intensity formula
for the $K$-allowed transitions
including the lowest-order corrections from Coriolis coupling.
In Ref.~[1], we also discuss the quantization for the $K$-forbidden
transitions and show explicit formulae for several applications to
the intra- and inter-band transitions.

The major achievement of this ``quantization''
is to provide a feasible method of microscopically
calculating the intrinsic parameters of
rotational intensity relations
including the effects of Coriolis coupling.
As an example of the applications,
we discuss here the Coriolis coupling effects for $I^\pi = 3^-$
octupole states.
Neerg\aa rd and Vogel~[3]
have shown that Coriolis mixing among the octupole states
is important even at low spins.
The relative differences of B($E3$; $3^- \rightarrow 0^+$) values
among the different $K$ modes cannot be understood
without Coriolis-coupling effects.
Fig.1 shows the results of the cranked RPA calculation
for the $E3$ amplitudes
associated with the octupole-vibrational bands ($K=0\sim 3$)
in Gd isotopes.
The Coriolis coupling concentrates the $E3$ strengths onto
the lowest-lying octupole state
($K=1$ in $^{156,158}$Gd and $K=2$ in $^{160}$Gd).
The large differences of B($E3$) values between the $K=0$ and 1 bands
observed in $^{156,158}$Gd disagrees with the leading-order calculation
without the Coriolis coupling (open circles),
while it is reproduced in the calculation with
the lowest-order correction terms (solid circles).
This example shows the usefulness of the new quantization method.

\vspace{-0.2cm}
\begin{flushleft}
{\bf References}

{\small
[1] Y.R.~Shimizu and T.~Nakatsukasa,
preprint {\sl nucl-th 960618}, to be published in
Nucl. Phys. A.

[2] E.R.~Marshalek, Nucl. Phys. {\bf A266} (1976) 317;
{\bf A275} (1977) 416.

[3] K.~Neerg\aa rd and P.~Vogel,
Nucl. Phys. {\bf A145} (1970) 33.
}

\begin{minipage}[t]{0.7\textwidth}
\vspace{-0.1cm}
\epsfxsize=\textwidth
\centerline{\epsfbox{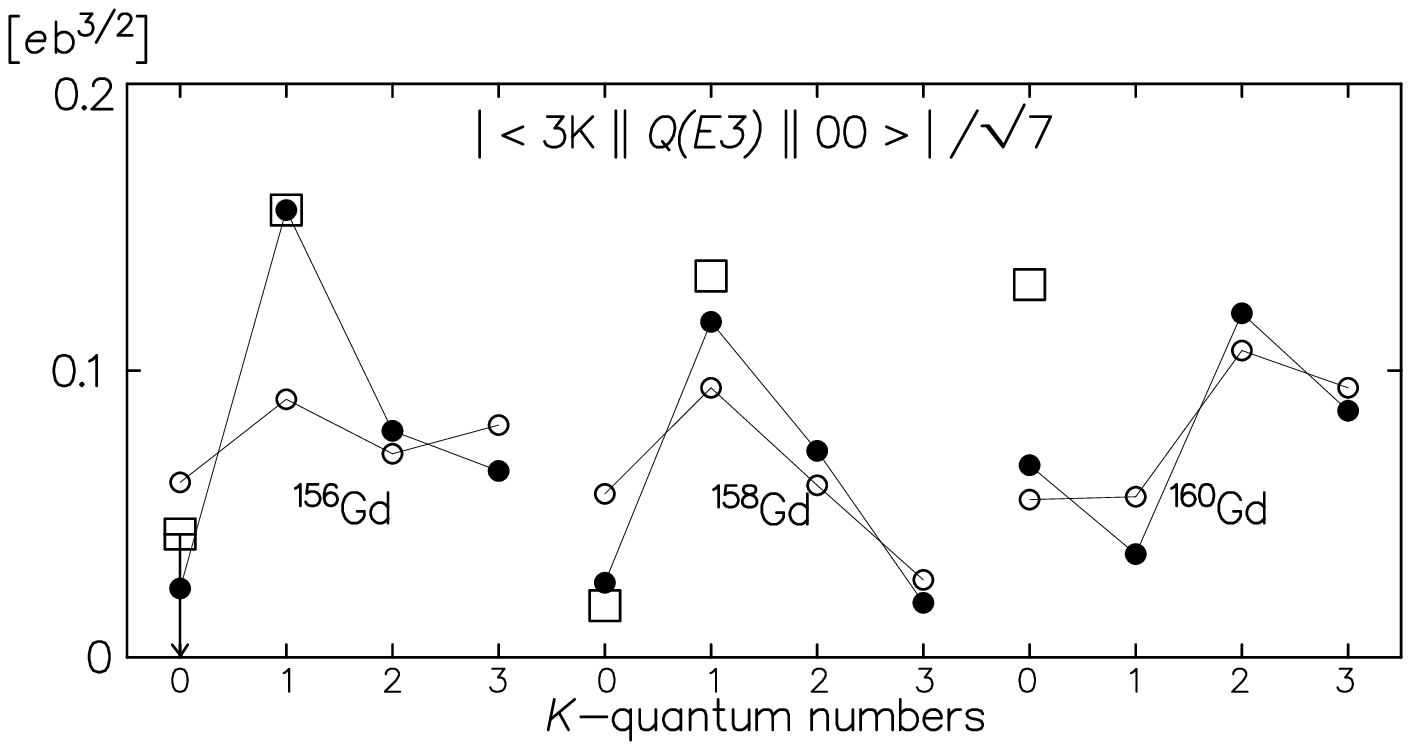}}
\end{minipage}
\begin{minipage}[t]{0.29\textwidth}
\vspace{1.4cm}
\small
{\bf Fig.1} :\ 
The $E3$ amplitudes for Gd isotopes
calculated by means of the RPA based on
the cranked shell model.
Solid (open) circles indicate the results with (without)
the lowest-order Coriolis coupling terms.
Experimental data are denoted by squares.
See Ref.~[1] for details of the calculations.
\end{minipage}

\end{flushleft}

\end{document}